\documentclass[twocolumn,showpacs,preprintnumbers,amsmath,amssymb]{revtex4}
\usepackage{graphicx}
\usepackage{dcolumn}
\begin{document}
\title{Search for extraterrestrial point sources of high energy neutrinos with AMANDA-II using data collected in 2000-2002}

\newcounter{foots}
\newcounter{notes}

\author{
M.~Ackermann$^{4}$,
J.~Ahrens$^{11}$, 
X.~Bai$^{1}$, 
R.~Bay$^{9}$,
M.~Bartelt$^{2}$,
S.~W.~Barwick$^{10}$, 
T.~Becka$^{11}$, 
K.~H.~Becker$^{2}$,
J.~K.~Becker$^{20}$, 
E.~Bernardini$^{4}$
% ...
\vspace*{0.2cm}
}

\affiliation{$^1$Bartol Research Institute, University of Delaware, Newark, DE 19716, USA}
\affiliation{$^2$Department of Physics, Bergische Universit\"at Wuppertal, D-42097 Wuppertal, Germany}
% ...
\affiliation{$^4$DESY, D-15735, Zeuthen, Germany}
% ...
\affiliation{$^9$Dept. of Physics, University of California, Berkeley, CA 94720, USA}
\affiliation{$^{10}$Dept. of Physics and Astronomy, University of California, Irvine, CA 92697, USA}
\affiliation{$^{11}$Institute of Physics, University of Mainz, Staudinger Weg 7, D-55099 Mainz, Germany}
% ...
\affiliation{$^{20}$Institute of Physics, University of Dortmund, D-44221 Dortmund, Germany}


\begin{abstract}

The results of a search for point sources of high energy neutrinos in the northern hemisphere using
data collected by AMANDA-II in the years 2000, 2001 and 2002 are presented. In particular, a
comparison with the single-year result previously published shows that the sensitivity was improved
by a factor of 2.2. The muon neutrino flux upper limits on selected candidate sources, corresponding to
an $E^{-2}$ neutrino energy spectrum, are included. Sky grids were used to search for possible
excesses above the background of cosmic ray induced atmospheric neutrinos. This search reveals no
statistically significant excess for the three years considered.

\end{abstract}

\maketitle

The detection of high energy cosmic rays raises fundamental questions about their generation and the
mechanisms responsible for such energies. The origin of cosmic rays above the ``knee"
($10^{15}\, \mathrm{eV}$) still remains uncertain. Nevertheless there is evidence that below such
energies they are generated through acceleration mechanisms in expanding supernova remnant shocks
\cite{snr} and in microquasars \cite{mquasar1,mquasar2,mquasar3}, although we cannot exclude the possibility of extragalactic
sources at these energies.
The interaction of accelerated protons with ambient matter or radiation leads to pion production and,
consequently, to neutrinos and gamma rays following a power-law energy spectrum.  
High energy gamma rays are affected by absorption during propagation and may also be produced by inverse
Compton scattering of shock-accelerated electrons. Therefore detection of gamma rays alone is insufficient evidence
for hadronic acceleration.
Neutrinos can provide a link to increased understanding of high energy cosmic rays, although
most predictions of high energy extraterrestrial neutrino fluxes conservatively require kilometer-scale detectors
\cite{gaisser, learned, halzen}. 

AMANDA-II \cite{amanda2} operates at the geographic south pole.
It is composed of 677 optical modules (OMs) -- photomultiplier tubes encased in glass pressure vessels --
spaced along 19 vertical cables (strings) arranged in concentric circles. The instrument spans a geometrical 
volume of clear glacial ice between depths of $1500$ and $2000$ m, with a diameter of $200$ m.
The AMANDA-II neutrino telescope has been in operation since 2000.
In this letter, we follow up on a previously published search for high-energy 
neutrino point sources from the data sample collected in 2000 \cite{2000ps}, using data from the
three years, 2000 to 2002. The sensitivity for the detection of point sources has significantly improved
in AMANDA, compared to 1997 \cite{1997ps} and 1999 results \cite{ecrs}, due to both detector performance and analyses
technique improvements.

A high-energy muon neutrino interacting with the ice or bedrock in the vicinity of the detector
produces a high-energy muon propagating a few kilometers when above 1 TeV. At these energies the mean
angular offset between the muon track and incident neutrino is less than $0.8^{\circ}$. The muon
track is reconstructed using the detection of
Cherenkov light emitted as it propagates through the array of OMs, and the likelihood of arrival time of
detected photons at each OM location. The resulting zenith-dependent median pointing resolution varies
between 1.5$^\circ$ and 2.5$^\circ$ \cite{reco}.

Muons are also produced by the interactions of cosmic rays in Earth's atmosphere. These {\it atmospheric muons} 
dominate the AMANDA-II trigger rate so the search for neutrino-induced muon tracks is only conducted in the northern
hemisphere, using Earth as an atmospheric muon filter. A second source of background is represented by atmospheric
muon tracks reconstructed as up-going. These events can be rejected using track quality criteria. The most important
source of background is the residual up-going flux of neutrinos produced in the atmosphere by the impact of cosmic
rays. These {\it atmospheric neutrinos} also serve as a verification of the detection principle and demonstrates
AMANDA's capability as a neutrino detector \cite{atmnu1,atmnu2}.
The search for possible extraterrestrial neutrinos begins with a dataset dominated by the well-understood atmospheric
neutrinos. This analysis selects a three-year sample of events with median energy of $\sim 1.3\, \mathrm{TeV}$ and
extending up to $\sim$ 100 TeV. Extraterrestrial neutrinos are beleived to be distinguished by a harder energy
spectrum, taken as proportional to $E^{-2}$ in this analysis.

The exposure of the present analysis is three times higher than that of the previous analysis \cite{2000ps}.
A different search strategy is used, which includes an explicit high energy event selection to reduce
the expected lower energy atmospheric neutrino background.

\section{Data Analysis}

The data used for this analysis were collected between the months of February and November in the years 
2000, 2001 and 2002 (see Table I).
\begin{table}[htb]
 \begin{center}
  \begin{tabular}{ccccc}

  \hline\hline
  year &            & livetime (days) &            & triggers            \\
  \hline
  2000 & $~~~~~~~~$ & 197             & $~~~~~~~~$ & $1.34\times 10^{9}$ \\
  2001 & $~~~~~~~~$ & 194             & $~~~~~~~~$ & $2.04\times 10^{9}$ \\
  2002 & $~~~~~~~~$ & 216             & $~~~~~~~~$ & $2.17\times 10^{9}$ \\
  \hline\hline
  \end{tabular}
  \label{tab:ystat}
  \caption{The experimental livetime and number of triggered events for each year used in this analysis.
		The triggered events may vary in different years mostly due to different cleaning procedures,
		which are mainly affected by the number of stable OMs during the specific year.
		}
 \end{center}
\end{table}  

The experimental sample used in this analysis corresponds to a total of 607 days of livetime and contains
almost 5.6 billion triggers.
Starting from 2002, a first level filter is performed at the South Pole during data taking. The reduced
amount of data is transferred via satellite to the northern hemisphere for analysis.
After the application of an iterative maximum-likelihood reconstruction algorithm and the selection of tracks that
are likely to be upgoing \cite{reco}, about 0.45 million events with reconstructed declination $\delta > -10^{\circ}$
remain. Since AMANDA-II is located at the south pole, $\delta = 0^{\circ}$ corresponds to horizontal and
$\delta = 90^{\circ}$ to vertical up-going directions.
These events, containing mostly mis-reconstructed atmospheric muons and a contribution of atmospheric
neutrinos, were used as an experimental background for selection optimization.

To avoid biasing the event selection the data were
scrambled by randomizing the reconstructed right ascension ($\alpha$) of each event. The
optimization procedure makes use of three observables: the number of hit OMs for each event ($\mathrm{nch}$),
the reconstructed track length in the array and the likelihood ratio between the muon track reconstruction and a muon
reconstruction constrained by using an atmospheric muon prior \cite{hill01}.
A full simulation chain, including neutrino absorption in the Earth, neutral current regeneration, muon
propagation and detector response for the given data taking periods, is used to simulate point sources
of muon neutrinos and anti-neutrinos \cite{2000ps}.
Events are simulated at the center of each
$5^{\circ}$ band of declination ($\delta$), according to an $E^{-2}$ energy spectrum.
The final cuts on these observables and the optimum size of each circular search bin were independently determined for
each declination band in order to have the strongest constraint on the signal hypothesis. This corresponds to
the best sensitivity, i.e. the average flux upper limit obtained in an ensemble of identical experiments assuming
no signal \cite{mrp}. The true directional information was then restored for the calculation of the limits.

The upper limits of this analysis were calculated using the background $n_{b}$ measured using the events
off-source in the corresponding declination band, and the expected number of events, $n_{s}$, from a simulated point source
of known flux $\Phi(E)$: $\Phi_{limit}(E) = \Phi(E) \times \mu_{90}(n_{obs}, n_{b})/n_{s}$. Here
$n_{obs}$ is the number of observed events in the given source bin, and $\mu_{90}$ is the upper limit
on the number of events following the unified ordering prescription of Feldman and Cousins \cite{fc98}.  
The three years were analyzed both separately and as combined data samples.

\begin{figure}[htb]
\begin{center}
\includegraphics[width=3.2in]{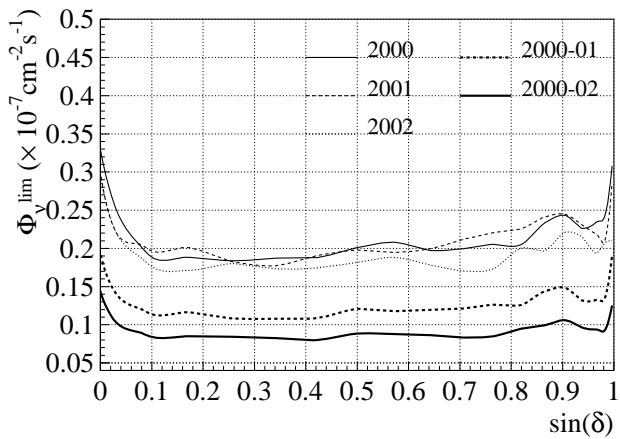}  
\end{center}
\vskip -0.6cm
\caption{Sensitivities on the integrated flux above $E_{\nu} = 10\, \mathrm{GeV}$ as a function of declination
	and for an $E^{-2}$ energy spectrum. The sensitivities for the year 2000, 2001 and 2002 are compatible with
	each other, and shown along with the one for 2000-01 and for the 2000-02 three-year sample.}
\label{fig:sens}
\end{figure}

\section{Calibration and Systematic Uncertainty}
 
Using the three-year experimental
sample the absolute normalization of the detector simulation, with respect to atmospheric neutrino flux,
was determined to be $1.03\pm 0.02$. This normalization factor is different from the value in reference \cite{2000ps},
but consistent with the theoretical uncertainty of 25\% \cite{lipari93}.
The optical properties of the fiducial ice, needed for the detector simulation, are determined using down-going
muon data and in-situ calibration lasers \cite{kurt}.
The overall experimental systematic uncertainty in the acceptance was evaluated to be $\sim$ 30\% \cite{down}.
The absolute pointing accuracy, determined with coincident
events between the SPASE air shower array \cite{spase} and AMANDA-II, is better than 1$^\circ$, i.e. smaller than
the angular resolution.
These systematic uncertainties do not affect significantly the limit calculations \cite{2000ps}.

\section{Results}

Figure \ref{fig:sens} shows the calculated sensitivity versus declination for energies above 10 GeV.
The event selection used produces a sensitivity which is fairly constant over all declinations.
For $0^{\circ} < \delta < 5^{\circ}$ (0 $<$ sin($\delta$) $<$ 0.09) the background contamination is 4 times higher than for
$\delta > 5^{\circ}$, and the sensitivity is poorer. For $\delta > 80^{\circ}$ (sin($\delta$) $>$ 0.98), on the
other hand, the solid angle aperture is small and the background evaluation is affected by higher relative
statistical fluctuations. The sensitivity improves with the detector exposure and for the three-year
sample it is $\sim$ 2.2 times better than for a single year. This improvement is better than would be
expected from longer exposure alone in the presence of background, due to improved analysis techniques.

\begin{table}[htb]
 \begin{center}
  \begin{tabular}{cccc}

  \hline\hline
  $~~~~$year$~~~~$   & $~~~~n_{obs}~~~~$ & $~~~~n_{p}(\nu_{\mu}^{atm})~~~~$ & $~~~~n_{p}(\nu_{\mu}^{sig})~~~~$ \\
  \hline
  2000\cite{2000ps}  & 601 & 676 & 133 \\
  \hline
  2000   &  306 & 296 & 111 \\
  2001   &  347 & 364 & 115 \\
  2002   &  429 & 429 & 131 \\
  \hline
  00-02  &  646 & 635 & 297 \\
  \hline\hline
  \end{tabular}
  \label{tab:res}
  \caption{The number of observed events with $\delta > 5^\circ$ after cut optimization, for each year and
	the combined three-year sample. The numbers relative to reference \cite{2000ps} are compatible with
	a normalization factor of $\sim 0.86$, for the atmospheric neutrino simulation, as quoted in the
	above reference. The numbers $n_p$ of the predicted atmospheric and signal neutrino events
	(with signal energy spectrum of
	${d\Phi_{\nu_{\mu}}\over dE}=10^{-6}\times E^{-2}\,\mathrm{cm}^{-2}\mathrm{s}^{-1}\mathrm{sr}^{-1}\mathrm{GeV}^{-1}$)
	are also shown.}
 \end{center}
\end{table}  

The final three-year sample consists of 646 upward ($\delta > 5^\circ$) reconstructed muons (see Table II).
The predicted number of atmospheric neutrinos is 635. In the year 2000 alone, the number of selected
events is 306, compared with the 601 (699 for $\delta > 0^\circ$) in reference \cite{2000ps}.
The difference between the two samples is due to the different choice of observables used for the selection
optimization. In particular the use of the $\mathrm{nch}$ observable, which is correlated to the energy released by the muon
in the array and, ultimately, to the neutrino energy, selects $\sim$ 26\% higher median energies than
those in \cite{2000ps} (from $\sim 700\, \mathrm{GeV}$ to $\sim 1\,\mathrm{TeV}$ for a single year).
This selection is obtained at the price of removing a significant
fraction of atmospheric neutrino events: for instance only 221 events in \cite{2000ps} would survive the new
selection, 94\% of which (i.e. 207) are also found in the new sample from the year 2000.

As shown in Table II the number of events in the final sample ($n_{obs}$) does not sum up with experimental exposure.
Since the signal hypothesis predicts a higher event intensity at high energy than the atmospheric neutrino
background, a longer exposure allows a stronger constraint on a given model by requiring a stronger energy cut
(the median energy increases from $\sim 1\,\mathrm{TeV}$ for a single year to $\sim 1.3\,\mathrm{TeV}$ for the three-year
sample), which rejects more background events and results in stricter limits.
Consequently the three-year sample contains $\sim$ 40\% fewer observed events than the sum of single years,
but only $\sim$ 17\% of the high energy neutrino signal events are lost.

\begin{figure}[htb]
\begin{center}
\includegraphics[width=3.2in]{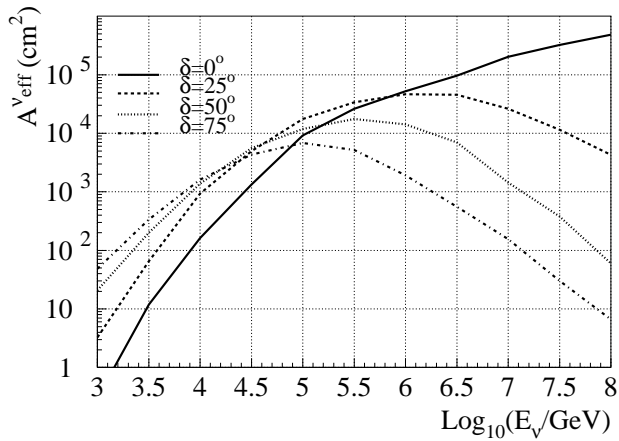}  
\end{center}
\vskip -0.6cm
\caption{Muon neutrino effective area as a function of the neutrino energy at different declinations. The effect of neutrino
         absorption in the Earth is responsible for the effective area decrease at high energies and declinations.}
\label{fig:aeff}
\end{figure}

The detector performance is assessed by the neutrino effective area $A^{\nu}_{eff}(E_{\nu},\delta)$, which contains the neutrino
interaction probability, muon propagation, detector response and the analysis selection. It is defined by the relation
between the differential neutrino flux ${d\Phi_{\nu}\over d\Omega dE_{\nu}}$ and the predicted number
of neutrino events $n_{p}(\nu)$, through the equation

\begin{equation}
n_{p}(\nu) = T_{live}\cdot\int_{\Omega}\int_{E_{\nu}^{min}}^{E_{\nu}^{max}} A^{\nu}_{eff}(E_{\nu},\delta){d\Phi_{\nu}\over d\Omega dE_{\nu}}d\Omega dE_{\nu}
\end{equation}

Figure \ref{fig:aeff} shows the muon neutrino effective area as a function of neutrino energy for the three-year
optimized selection. The curves are shown for different declinations. Above 1 $\mathrm{PeV}$
neutrinos begin to be absorbed by the Earth, except for the events that enter AMANDA-II horizontally.

A binned search for excesses in the $5^{\circ}<\delta<85^{\circ}$ region was performed on the three-year
event sample. The search grid contains 290 rectangular bins with declination-dependent width ranging from
$5.6^{\circ}$ to $8.8^{\circ}$, based on the optimized search bin diameter. The grid is shifted 4 times in $\delta$ and
$\alpha$ to fully cover boundaries between the bins of the original configuration. 
A higher number of grid shifts showed no improvement in the average maximum statistical significances on simulated
Poisson-fluctuated signal with intensities comparable to the background.
The probability distribution for background fluctuations in the ensemble of bins was evaluated by
using 20,000 experimental samples with scrambled $\alpha$ and calculating the highest
value of the maximum statistical fluctuation significance over the entire sky.

\begin{table}[htb]
 \begin{center}
  \begin{tabular}{lcccccccc}\hline\hline

 Candidate & & & from & \cite{2000ps} & & this & work \\ \hline
           & $\delta$($^\circ$) & $\alpha$(h) 	&
		$n_{\mathrm{obs}}$ & $n_{b}$ 	&
		$\Phi_{\nu}^{\mathrm{lim}}$ 	&
		$n_{\mathrm{obs}}$ & $n_{b}$	&
		$\Phi_{\nu}^{\mathrm{lim}}$ \\ \hline
   \multicolumn{7}{c}{ \emph{TeV Blazars} } \\
      {\footnotesize Markarian 421 } & 38.2 & 11.07 & 3 & 1.50 & 3.5 & 0 & 1.35 & 0.34 \\
      {\footnotesize Markarian 501 } & 39.8 & 16.90 & 1 & 1.57 & 1.8 & 3 & 1.31 & 1.49\\
      {\footnotesize 1ES 1426+428  } & 42.7 & 14.48 & 1 & 1.62 & 1.7 & 2 & 1.13 & 1.16\\
      {\footnotesize 1ES 2344+514  } & 51.7 & 23.78 & 1 & 1.23 & 2.0 & 1 & 1.25 & 0.82\\
      {\footnotesize 1ES 1959+650  } & 65.1 & 20.00 & 0 & 0.93 & 1.3 & 0 & 1.59 & 0.38\\  \hline
   \multicolumn{7}{c}{ \emph{GeV Blazars} } \\
      {\footnotesize QSO 0528+134  } & 13.4 &  5.52 & 1 & 1.09 & 2.0 & 1 & 1.88 & 0.57\\
      {\footnotesize QSO 0235+164  } & 16.6 &  2.62 & 1 & 1.49 & 1.7 & 3 & 2.15 & 1.12\\
      {\footnotesize QSO 1611+343  } & 34.4 & 16.24 & 0 & 1.29 & 0.8 & 0 & 1.66 & 0.31\\
      {\footnotesize QSO 1633+382  } & 38.2 & 16.59 & 1 & 1.50 & 1.7 & 1 & 1.33 & 0.75\\
      {\footnotesize QSO 0219+428  } & 42.9 &  2.38 & 1 & 1.63 & 1.6 & 0 & 1.15 & 0.37\\
      {\footnotesize QSO 0954+556  } & 55.0 &  9.87 & 1 & 1.66 & 1.7 & 2 & 1.04 & 1.50\\
      {\footnotesize QSO 0716+714  } & 71.3 &  7.36 & 2 & 0.74 & 4.4 & 3 & 0.93 & 1.91\\ \hline
   \multicolumn{7}{c}{ \emph{Microquasars} } \\
      {\footnotesize SS433         } &  5.0 & 19.20 & 0 & 2.38 & 0.7 & 1 & 2.21 & 0.55\\
      {\footnotesize GRS 1915+105  } & 10.9 & 19.25 & 1 & 0.91 & 2.2 & 3 & 1.84 & 1.26\\
      {\footnotesize GRO J0422+32  } & 32.9 &  4.36 & 2 & 1.31 & 2.9 & 2 & 1.49 & 1.08\\
      {\footnotesize Cygnus X1     } & 35.2 & 19.97 & 2 & 1.34 & 2.5 & 0 & 1.59 & 0.31\\
      {\footnotesize Cygnus X3     } & 41.0 & 20.54 & 3 & 1.69 & 3.5 & 1 & 1.26 & 0.75\\
      {\footnotesize XTE J1118+480 } & 48.0 & 11.30 & 1 & 0.92 & 2.2 & 1 & 1.12 & 0.80\\
      {\footnotesize CI Cam        } & 56.0 &  4.33 & 0 & 1.72 & 0.8 & 2 & 1.05 & 1.44\\
      {\footnotesize LS I +61 303  } & 61.2 &  2.68 & 0 & 0.75 & 1.5 & 5 & 1.67 & 2.43\\  \hline
   \multicolumn{7}{c}{ \emph{SNR, magnetars \& miscellaneous} } \\
      {\footnotesize SGR 1900+14   } &  9.3 & 19.12 & 0 & 0.97 & 1.0 & 2 & 1.78 & 0.94\\
      {\footnotesize Crab Nebula   } & 22.0 &  5.58 & 2 & 1.76 & 2.4 & 4 & 1.86 & 1.43\\
      {\footnotesize Cassiopeia A  } & 58.8 & 23.39 & 0 & 1.01 & 1.2 & 2 & 1.12 & 1.38\\
      {\footnotesize 3EG J0450+1105} & 11.4 &  4.82 & 2 & 0.89 & 3.2 & 1 & 1.83 & 0.59\\
      {\footnotesize M 87          } & 12.4 & 12.51 & 0 & 0.95 & 1.0 & 3 & 1.83 & 1.24\\
      {\footnotesize Geminga       } & 17.9 &  6.57 & 3 & 1.78 & 3.3 & 2 & 2.06 & 0.81\\
      {\footnotesize UHE CR Triplet} & 20.4 &  1.28 & 2 & 1.84 & 2.3 & 0 & 2.15 & 0.20\\
      {\footnotesize NGC 1275      } & 41.5 &  3.33 & 1 & 1.72 & 1.6 & 1 & 1.14 & 0.78\\
      {\footnotesize Cyg. OB2 region.} & 41.5 & 20.54 & 3 & 1.72 & 3.5 & 1 & 1.14 & 0.78\\
      {\footnotesize UHE CR Triplet} & 56.9 & 12.32 & 1 & 1.48 & 1.9 & 1 & 1.17 & 0.93\\

 \hline\hline

  \end{tabular}
  \label{tab:limits}
  \caption{90\% CL upper limits on candidate sources. Results from the present analysis are reported for a
		comparison with the limits from \cite{2000ps}. Limits are for the assumed $E_{\nu}^{-2}$
		spectral shape, integrated above
		$E_{\nu} = 10\, \mathrm{GeV}$, and in units of $10^{-8} \mathrm{cm}^{-2} \, \mathrm{s}^{-1}\, (\Phi_{\nu}^{\mathrm{lim}})$.}
 \end{center}
\end{table}

The bin with the most statistically significant excess from the three-year experimental sample is at about
$\alpha=22h$ and $\delta=21^{\circ}$,
with 10 observed events in the search bin on a background of 2.38 events, estimated from the corresponding declination band.
The observed excess has a statistical significance of 1.9$\times 10^{-4}$ (3.73 $\sigma$). The chance probability of
such an excess, in the ensamble of bins, is 28\%.

Table III shows the 90\% CL neutrino flux limits for northern hemisphere TeV blazars, selected GeV blazars, microquasars,
magnetars and selected miscellaneous candidates. The limits are compared with the values from \cite{2000ps}, they
are compatible with the average flux upper limit, or sensitivity, of Figure \ref{fig:sens} and the deviations from it are
due to statistical fluctuations in the observed sample.

\begin{figure}[htb]
\begin{center}
\includegraphics[width=3.4in]{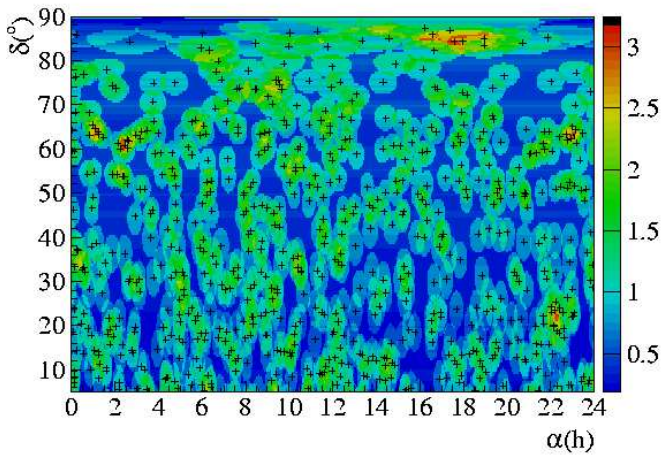}  
\end{center}
\vskip -0.6cm
\caption{2000-02 upper limits (90\% CL) on the neutrino flux integrated above 10 GeV in equatorial
	coordinates for $\delta > 5^{\circ}$. Limits (scale on right axis) are given in units of
	$\times 10^{-8}$ cm$^{-2}$s$^{-1}$ for the assumed $E^{-2}$ spectrum. Systematic uncertainties
	are not included. The cross symbols represent the observed events.}
\label{fig:limits}
\end{figure}

Figure \ref{fig:limits} shows the 90\% CL neutrino flux upper limits in equatorial coordinates.
The limits are calculated by scanning the sky and counting the events within the optimized search bins
at the given declination. The highest upper limit in the Figure corresponds to the previously discussed statistically
significant bin. Other high limit spots visible in the figure have statistical significances smaller
than 3.4 $\sigma$. 

We analyzed the 2000-02 data sample collected by the AMANDA-II detector to search for point sources of high
energy neutrinos. We performed both a non-targeted binned search and a targeted search focussing on
known objects that are potential high energy neutrino emitters (as in reference \cite{2000ps}). We found no
evidence of a significant flux excess above the background. A km-scale experiment, such as IceCube \cite{icecube},
will be able to increase the detection sensitivity by at least a factor of 30 in the same time scale.

\section{Acknowledgments}

{We acknowledge the support of the following agencies: National
Science Foundation--Office of Polar Programs, National Science
Foundation--Physics Division, University of Wisconsin Alumni Research
Foundation, Department of Energy, and National Energy Research
Scientific Computing Center (supported by the Office of Energy
Research of the Department of Energy), UC-Irvine AENEAS Supercomputer
Facility, USA; Swedish Research Council, Swedish Polar Research
Secretariat, and Knut and Alice Wallenberg Foundation, Sweden; German
Ministry for Education and Research, Deutsche Forschungsgemeinschaft
(DFG), Germany; Fund for Scientific Research (FNRS-FWO), Flanders
Institute to encourage scientific and technological research in
industry (IWT), and Belgian Federal Office for Scientific, Technical
and Cultural affairs (OSTC), Belgium; I.T. acknowledges support from 
Fundaci\'{o}n Venezolana de Promoci\'{o}n al Investigador (FVPI), 
Venezuela;  D.F.C. acknowledges the support of the NSF CAREER program.}


\begin{thebibliography}{99}

\bibitem{snr}
F.~A.~Aharonian {\em et al.}, Nature {\bf 432}, 75 (2004).

\bibitem{mquasar1}
M.~Massi {\em et al.} (2004), astro-ph/0410504.

\bibitem{mquasar2}
V.~Bosch-Ramon and J.M. Paredes (2004), astro-ph/0401260.

\bibitem{mquasar3}
C.~Distefano, D.~Guetta, A.~Levinson and E.~Waxmann, Astrophys. J. {\bf 575}, 378 (2002).

\bibitem{gaisser}
T.~K.~Gaisser, F.~Halzen and T.~Stanev, Phys. Rept. {\bf 258}, 173 (1995).

\bibitem{learned}
J.~G.~Learned and K.~Mannheim, Ann. Rev. Nucl. Part. Sci. {\bf 50}, 679 (2000).

\bibitem{halzen}
F.~Halzen and D.~Hooper, Rept. Prog. Phys. {\bf 65}, 1025 (2002).

\bibitem{amanda2} 
E.~Andr\'{e}s, {\em et al.}, Astropart. Phys. {\bf 13}, 1 (2000).

\bibitem{2000ps}
J.~Ahrens {\em et al.}, Phys. Rev. Lett. {\bf 92}, 071102 (2004).

\bibitem{1997ps}
J.~Ahrens {\em et al.}, Astrophys. J. {\bf 583}, 1040 (2003).

\bibitem{ecrs}
P.~Desiati {\em et al.}, in {\em Proceedings of the 19$^{th}$ European Cosmic Ray Symposium, Florence, Italy}
(2004), to be published in Int. J. Mod. Phys. A.

\bibitem{reco}
J.~Ahrens {\em et al.}, Nucl. Inst. Meth. {\bf A524}, 169 (2004).

\bibitem{atmnu1}
E.~Andr\'{e}s, {\em et al.}, Nature {\bf 410}, 441 (2001)

\bibitem{atmnu2}
J.~Ahrens {\em et al.}, Phys. Rev. {\bf D66}, 012005 (2002)

\bibitem{hill01} 
G.~C.~Hill, in {\em Proceedings of the 27$^{th}$ ICRC, Hamburg, Germany}, {\bf HE 267}, 1279,
edited by K.-H. Kampert, G. Hainzelmann and C. Spiering, Copernicus Gesellschaft e.V.,
Katlemburg-Lindau, Germany, vol. HE 267, p. 1279.

\bibitem{mrp}
G.~C.~Hill and K.~Rawlins, Astropart. Phys. {\bf 19}, 393 (2003).

\bibitem{fc98} 
G.~J.~Feldman and R.~D.~Cousins, Phys. Rev. {\bf D57}, 3873 (1998).

\bibitem{lipari93} P.~Lipari, Astropart. Phys. {\bf 1}, 195 (1993).

\bibitem{kurt}
K.~Woschnagg {\em et al.}, in {\em Proceedings of the 21$^{st}$ International Conference On Neutrino Physics
And Astrophysics (Neutrino 2004), Paris, France} (2004), to be published in Nucl. Phys. B (Proceedings
Supplement), astro-ph/0409423.

\bibitem{down}
P.~Desiati {\em et al.}, in {\em Proceedings of the 28$^{th}$ ICRC, Tsukuba, Japan},
edited by T. Kajita, Y. Asaoka, A. Kawachi, Y. Matsubara, M. Sasaki (Univ. Acad. Pr., Tokyo, 2003),
vol. HE 2.3, p. 1373.

\bibitem{spase}
J.~E.~Dickinson {\em et al.}, Nucl. Inst. Meth. {\bf A440}, 95 (2000).

\bibitem{icecube}
J.~Ahrens {\em et al.}, Astropart. Phys. {\bf 20}, 507 (2004).

\end{thebibliography}

\end{document}